# A Photonic Atom Probe coupling 3D Atomic Scale Analysis with *in situ* Photoluminescence Spectroscopy

J. Houard[1], A. Normand[1], E. Di Russo[1], C. Bacchi[1], P. Dalapati[1], G. Beainy[1], S. Moldovan[1], G. Da Costa[1], F. Delaroche[1], C. Vaudolon[1], J.M. Chauveau[2], M. Hugues[2], D. Blavette[1], B. Deconihout[1], A. Vella[1], F. Vurpillot[1] and L. Rigutti[1,*]

1) UNIROUEN, CNRS, Groupe de Physique des Matériaux, Normandie Université, 76000 Rouen, France.

2) Centre de Recherche sur l'Hétéro-Epitaxie et ses Applications, UPR10 CNRS, 06560 Valbonne, France.

**Abstract**

Laser enhanced field evaporation of surface atoms in Laser-assisted Atom Probe Tomography (La-APT) can simultaneously excite phtotoluminescence in semiconductor or insulating specimens. An atom probe equipped with appropriate focalization and collection optics has been coupled with an in-situ micro-Photoluminescence (µPL) bench that can be operated during APT analysis. The Photonic Atom Probe instrument we have developed operates at frequencies up to 500 kHz and is controlled by 150 fs laser pulses tunable in energy in a large spectral range (spanning from deep UV to near IR). Micro-PL spectroscopy is performed using a 320 mm focal length spectrometer equipped with a CCD camera for time-integrated and with a streak camera for time-resolved acquisitions. An exemple of application of this instrument on a multi-quantum well oxide heterostructure sample illustrates the potential of this new generation of tomographic atom probe.

* Corresponding author: lorenzo.rigutti@univ-rouen.fr

# 1. Introduction

Correlative microscopy is crucial in nanoscience and nanotechnology, because they make it possible to establish the relationship between structural and functional properties of systems containing a nanometer-scale functional element, or even of individual nanometric objects. Correlative studies can be implemented at different levels.

(i) Statistical correlations can be established by applying different techniques to the study of different nanometric parts of the same macroscopic sample [1].

(ii) A sequential correlation, or *multi-microscopy*, can generally be obtained by sequentially applying different techniques on the same object at the nanometric scale [2].

(iii) In situ correlation of the properties of individual nano-objects can also be achieved by applying in situ techniques. Cathodoluminescence is a significant example of technique coupling structural analysis by electron microscopy with optical analysis [3].

Laser-assisted Atom Probe Tomography (La-APT) has emerged as a technique allowing for the reconstruction in 3D of the chemical composition of a nanoscale specimen with a precision close to the atomic scale [4]. Typical specimens are prepared in the form of needles with an apex radius of about 50 nm. Up to date, the technique has been successfully applied to a large variety of semiconductor and dielectric materials, including those on which technologically relevant systems such as transistors or light-emitting diodes are based [1], [5]–[8]. Various strategies have been recently applied to the improvement of the technique performances. A part of the effort has been devoted to instrumental development, including the introduction of high-energy light sources for specimen pulsing [9] or the coupling of APT with electron imaging and diffraction for a dynamic control of tip shape evolution during the analysis [10].

In past studies our laboratory has shown that the study of micro-photoluminescence (µPL) emitted from samples under the form of nanoscale APT specimens yields more detailed information than under the form of larger-scale objects (thin films or bulk, for instance) [2], [11]. Besides the original information provided by this correlative microscopy approach, focused laser pulses used to field evaporate ions can simultaneously be used to generate a µPL signal that can be exploited during APT analysis. This allows not only a correlation at the nanometric scale of the structural and optical properties along the depth of the analyzed specimen, but also to study the effect of the change of the structure on the optical emission during evaporation.

Tomographic Atom Probe (TAP) relies on the generation of a high static electric field (20 - 50 V/nm) at the tip surface that promote surface atoms to evaporate as ions. This electric field is at the origin of a mechanical tensile stress of up to 10 GPa. This stress represents a critical issue for a number of APT analyses, as it is at the origin of moderate to severe plastic deformation within the specimen, leading to dislocation motion [12] up to specimen fracture. The study of the PL from color centers contained in diamond nanoneedles submitted to high electric field has recently allowed the optical measurement of this stress [13].

In the present paper, we show the potential and performance of this new generation of instrument combining µPL and atom probe tomography. We refer to this instrument as a Photonic Atom Probe (PAP). µPL studies can either be performed from non-evaporating or from evaporating specimens. This second option also allows consider µPL in the PAP as an *operando* spectroscopy technique. We show that this correlative instrument can operate either in a time-integrated or in a time-resolved mode. We will describe the main technical features of this instrument and illustrate its potential through the analysis of oxide heterostructures containing localized emitters such as quantum wells.

The potential applications of this instrument can be classified into two main domains [14], namely correlative microscopy and nanoscale physics.

In the field of correlative microscopy, the "natural" framework would be the study of the relationship between morphological, chemical and optical properties of nanoscale systems. The latter can be quantum dots, quantum wells, semiconductor alloys, extended or point defects exhibiting specific PL signatures. This approach is straightforwardly applicable to a large number of semiconductor systems provided that they produce sufficiently intense luminescence. This condition will be fulfilled using large enough time intervals. However, a large time interval will lead to the evaporation of a large fraction of the specimen. Hence, a compromise has to found so as to optimize the depth resolution of measurements.

The recorded µPL spectrum may be directly linked to the composition of the material. This will make it possible to check whether the composition measured by APT is correct. Additionally, the PL signal will evolve during field evaporation giving information on the in-depth evolution of the composition of the samples analyzed during the specimen evaporation. This information can be correlated with the structural information (in three dimensions) obtained by APT.

The second domain of application of our new instrument is nanoscale physics related to light-matter interaction under high electric field or high tensile stress in a needle-shaped specimen. Original insight can be obtained through the study of the dependence of optical signatures on the environmental parameters that can influence ion evaporation in APT. In particular, it will become possible to monitor the stress/strain state at the position occupied by a light emitter within an APT specimen. This is possible by recording the influence of stress/strain on the optical signature of the emitter itself as a function of the applied bias or during the evaporation of the specimen. It can also be expected that the variation of the PL intensity over time could be put in relationship with the laser absorption within the specimen either as a function of the applied bias or while its shape evolves during APT analysis.

## 2. Instrumental details

A schematic representation of the instrument is depicted in Fig. 1. It consists of a home-made La-APT chamber. The laser source controlling field evaporation is directed onto the electrically polarizable specimen through a focusing element. The laser provides the required energy to excite both field evaporation of surface atoms and electronic transitions giving rise to PL. During APT operation, ions accelerated by the electric field move towards the detector, while the PL signal is collected through the same path used for focusing the laser. Alternative paths to this back-scattering configuration, specific of our setup, may be implemented if needed. The PL signal is then analyzed by a spectrometer. APT and µPL can be operated simultaneously or independently. In the following subsections, the different parts of the instrument will be described.

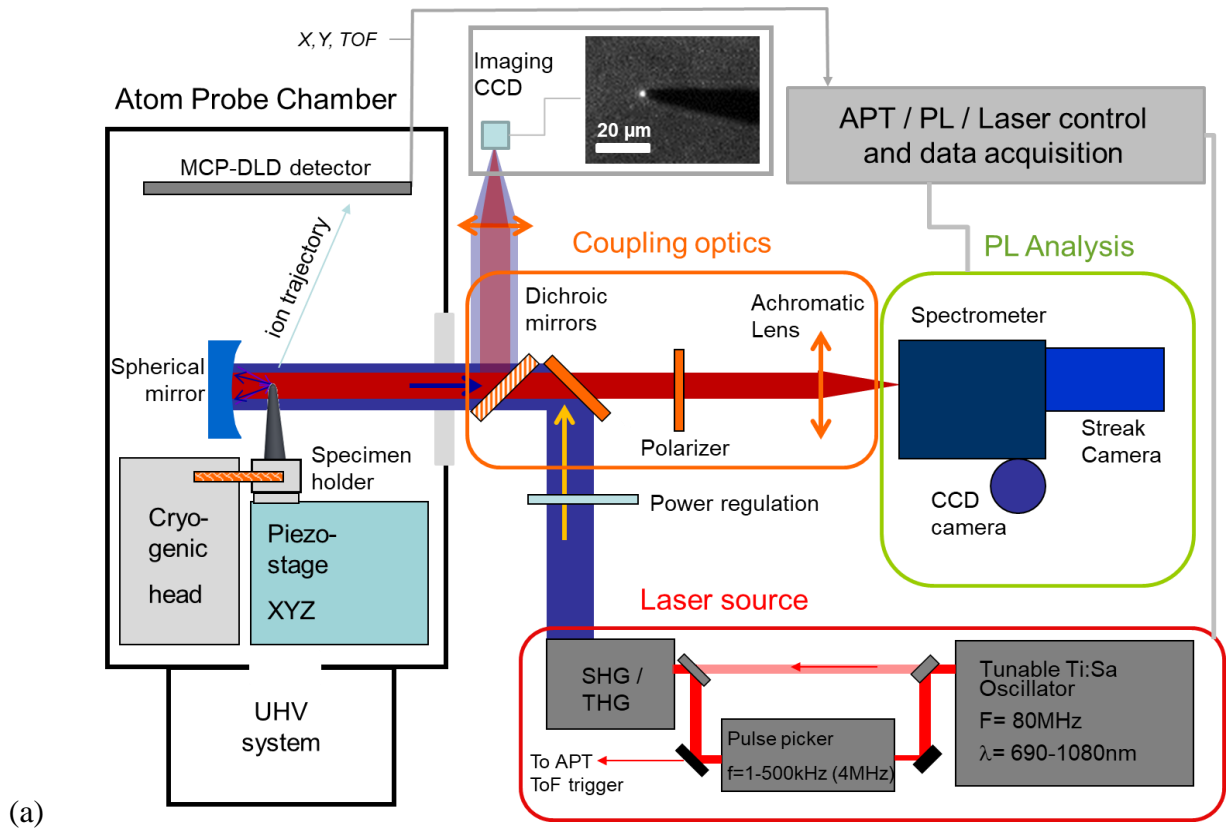

(a)

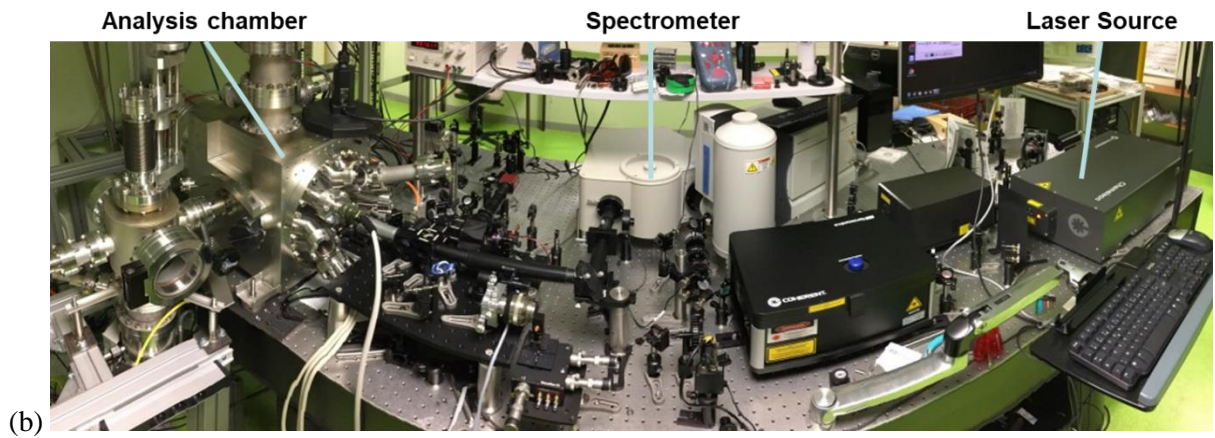

(b)

Fig. 1. (a) Schematics of the coupled µPL-APT instrument. (b) Panoramic picture of the experimental setup

### 2.1. Laser source

The laser source is a Ti-Sa tunable pulsed laser (a Chameleon Ultra II from Coherent). The pulse width is 150 fs, while the wavelength of the emitted photons can be tuned in the range 600 - 1100 nm at a frequency of 80 MHz with a peak power of 4W (or 50 nJ per pulse) for emission at 780 nm. The pulse frequency is divided by a pulse picker (an acousto-optic SiO2 Bragg cell from Coherent with a contrast ratio of 1000:1 and an efficiency of 50 %), so that the time interval between subsequent pulses is longer than the time of flight of the heaviest ions evaporated from the specimen. The laser frequency used during APT operation is typically 400 kHz. This value is set considering the typical ion times of flight (ToF) of evaporated ions in APT: the laser pulse triggering the evaporation also triggers the starting point for the measurement of the ToF, so the detection event would not be correctly identified if it occurs after a second ToF-triggering pulse. However, the laser frequency may be set to a higher value (up to 4 MHz) if the PL measurement is performed in static conditions (i.e. without evaporation) in order to enhance the PL signal intensity. After the pulse picker, the laser light can be directed into a second – and third-harmonic generator (HarmoniXX THG provided by APE that generates SHG and THG with walk off and delay compensation; generation efficiency is 20 % for SHG and 10 % for THG) in order to obtain light in the 300 - 550 nm and in the 200 - 366 nm. A fine regulation of the laser power is performed (with a half-wave plate and an alpha-BBO Glan-Laser polarizer) and a band pass filter if finally used (to remove all unwanted wavelength) before the beam enters the optical path coupling it into the APT chamber.

### 2.2. Optics

The present instrument implements free-space optics for the coupling of the laser with the analyzed specimen and for the collection of the light outside the APT chamber. The optical path is that of a standard µPL experiment. The laser is transmitted to the APT chamber through a UV to near-IR transparent fused silica UHV viewport and is directed onto the specimen through a dichroic mirror adapted to the reflection of the chosen laser wavelength and focused through a spherical mirror of numerical aperture NA = 0.3 and working distance WD = 25 mm. It should be noticed that the collimated beam illuminates the tip specimen at low intensity before being focused onto its apex by the spherical mirror. Because of the shape of needle-samples (a few mm long for 0.1 mm of diameter) compare to the size of the collimated laser, only a small fraction of the light is lost (< 5 – 10 %) and it affects the focal point only a little. Although the diffraction limit for this configuration sets a minimal beam size of $w_{diff} = \lambda/(2NA) \sim 450$ nm for laser excitation at $\lambda = 266$ nm, the effective laser spot size has been estimated to be rather of the order of $w_{eff} = 1.5 - 2$ µm. These limitations mainly come from spherical aberration of the focusing mirror. Far from being a disadvantage, this larger spot size makes the system more tolerant with respect to specimen drifts and fluctuations, provided they remain limited to a scale of around one tenth of the beam waist size. The collection of PL is performed through the same optical path in a backscattering configuration. A part of the beam emerging from the APT chamber can be redirected on a CCD imaging camera, which is used for coarse tip alignment and PL spot visualization. The image is formed by a UV-corrected triplet with 180 mm focal length that give us a magnification of ~7. From the pixel size of the CCD camera, details with size under 1 µm can be imaged. The PL signal is transmitted through the dichroic mirror, may optionally pass through a linear polarizer if polarization-resolved PL is wished and is focused at the entrance slit of the spectrometer by a UV corrected triplet with a focal length of 135 mm, a chromatic shift of 0.6 mm from 400 to 1000 nm and NA = 0.08 (that correspond to the numeric aperture of our spectrometer).

It can be mentioned that other coupling schemes may be implemented. It is indeed possible, in principle, to excite and collect through parabolic mirrors, to decouple the excitation and the collection path, and also to adopt a fiber optics coupling. These possibilities are beyond the scope of this work and will not be discussed in detail.

## 2.3. Atom Probe Chamber

The APT chamber hosts a straight flight path ion optics. The main elements of the APT chamber are described in the following subsections.

### 2.3.1. Vacuum and airlock

The main chamber is a cubic custom design one made of 304L stainless steel and every flanges are sealed with ConFlat copper gaskets. Four large flanges of 250 and 300 mm allow to easily place pumps, APT detector and optical bench. Many 25- and 40-mm flanges are available for sample transfer, viewport all needed equipment. Finally the top flange of 150 mm is used for cryogenic cooling feedthrough.

The chamber is pumped by a large turbomolecular hybrid bearing pump with a pumping speed of 1200 L/s for nitrogen (Hipace 1200 from Pfeiffer Vacuum). For good optical stability, the main chamber lays on the optical table, so that the turbomolecular pump is horizontally positioned on the back flange. The primary vacuum needed by the turbomolecular pump is done by a dry pump (ACP 15 from Adixen). Ultimate vacuum is improved by a NEG pump with a pumping capacity of 400 L/s for hydrogen (CapaciTorr D400 from SAES). The overall can be backed at 150 °C and the typical vacuum measured by a cold cathode gauge is $4 \cdot 10^{-9}$ Pa (IKR 070 from Pfeiffer Vacuum).

Samples are introduced in the main chamber with a magnetic passthrough after being pumped down to $10^{-6}$ Pa in an airlock. The airlock is equipped with a turbomolecular pump (a 210L/s from Pfeiffer Vacuum) and separated from the main chamber by a pneumatic valve (Mini UHV Gate from VAT).

### 2.3.2. Cryogenics and specimen holder

A fundamental aspect of the instrument, needing a dedicated technical solution with respect to standard APT systems, concerns the positioning and stabilization of the specimen at cryogenic temperatures. An optical bench has been built in the main chamber. It is composed of a 10 mm thick and 200 mm diameter 304L stainless steel plate fixed on the bottom flange of the chamber. On it, the specimen holder and the focusing spherical mirror are fixed. To have them together on the same base limits the drift.

The specimen holder system has a two-stage base. The lower stage accommodates a three-axis piezo-inertial stage (Smaract brand, composed of SLC 1740 linear axis controlled by a MSC-3CC controller offering 25 mm travel range in the three axis with a 2 nm resolution) which allows to very precisely set and control the specimen position.

The upper stage of the specimen holder is an oxygen free copper piece containing a highly thermal conductive but electrically isolated sapphire ceramic part. At the middle of it, the sample is fixed. The aim is the be able to put high voltage on the sample and to cool it down to cryogenic temperature. Above the sample holder, the liquid helium closed circuit cryogenic head (a CS 204-DMX-20-OM from ARS company) is connected to the specimen holder. As a specific feature of the present instrument, the cryogenic head is not in direct contact but, on the contrary, it is distant from the sample holder. It is indeed crucial that the thermal transfer is performed without transmission or almost without transmission of mechanical stress and/or oscillatory movements and/or vibrations. The thermal transfer is thus provided by copper braids connecting the cryogenic head and the specimen holder. The braids also allow for the specimen motion using the piezo stage. The upper stage and the lower stage are fixed with a PEEK (polyetheretherketone) part to thermally isolate the cryogenic part from the piezo stage that remains at room temperature. Thus, the mechanical effects of the cryogenic head are minimized or even eliminated by the use of the aforementioned transfer means which, at the same time, block the transmission of mechanical stresses from the cryogenic head to the sample support and provide thermal conduction between them, making it possible to reach the temperatures in interval 20 – 80 K favorable for tomography or even photoluminescence analyzes.

This allows for long analyses (over several hours even over an entire day) while keeping the position of the laser beam on the same part of the specimen, the center of the beam being centered with an accuracy of the order of a hundred nanometers, without any need for re-centering the specimen as in current commercial APT systems.

### 2.3.3. APT Detector and data acquisition interface

The APT analysis relies on a Multi-Channel Plate Delay Line Detector (MCP-DLD) [15]. In the present case, the 77 mm diameter MCP is set at an approximate distance L = 18.5 cm from the specimen, roughly corresponding to a physical field of view of 11.7° (the effective field of view should be found by multiplying the physical field of view by the $m$+1~1.6 image compression factor). The MCP and the DLD signals (the X and Y coordinates of the impact on the detector surface, as well as the time of flight ToF of the ion) are collected and processed through a Keysight U1051A Acquiris board. This system allows for a sufficiently high acquisition speed, compatible with APT operation at a laser frequency as high as 500 kHz. Tests performed on the systems indicate that APT could be run at 1.7 MHz with a detection rate of 1 ion/pulse without any loss of information caused by memory management. However, in the current prototype the timing information is retrieved by the pulse signals through a simple discriminator, while in common APT systems a so-called advanced-DLD [16] is implemented, based on the use of a constant fraction discriminator, which reduces the impact of potential timing errors. Furthermore, the current system would have a reduced performance in multiple event detection, as two overlapping peaks would be treated as a single peak if the signal amplitude does not fall below a given threshold between the two peaks. This may induce significant biases due to the detection losses of multiple events, which could, in turn, affect the measured composition. However, these limitations may be technically overcome in a future instrument upgrade, through the implementation of a MCP-aDLD detector [16].

### 2.4. Photoluminescence Spectroscopy

The PL signal is focused onto the entrance slit of a Horiba iHR 320 grating spectrometer with 320 mm focal length. It is equipped with three different gratings, with densities of 150, 600 and 1200 lines/mm yielding the highest spectral resolution of around 0.5 meV and the largest spectral interval for CCD acquisition of 550 nm. The spectrometer is equipped with a liquid nitrogen-cooled CCD camera, which is used for time-integrated acquisitions in the range 200-1100 nm. Alternatively, the signal can be detected with a streak camera, which is applied to time-resolved acquisitions [17]. The streak camera photocathode is sensitive to photons in the 200 - 900 nm interval. Spectrally-resolved luminescence transients with time scales of 2 - 100 ns interval with an ultimate time resolution of 25 ps in a streaking range of 2 ns can be recorded.

3. **Study of an oxide heterostructure system by simultaneous time-integrated PL and APT analysis**

Simultaneous APT and µPL analyses have been conducted on a specimen composed of 20 ZnO/(Mg,Zn)O Quantum Wells (QWs). The structural, chemical and optical properties of this system have been thoroughly addressed by sequential correlative microscopy in a previous study, to which the reader is referred for details [18]. In Fig. 2-(i) we report a High-Angle Annular Dark Field (HAADF) Scanning Transmission Electron Microscopy (STEM) micrograph of the system. The ZnO QWs correspond to the bright contrast layers, with a thickness of around 2.9 nm, within the approximately 6.8 nm thick darker regions, which correspond to the (Mg,Zn)O barriers. All interfaces are not flat, but they exhibit a V-groove morphology which is known to propagate from the substrate to the epitaxial layers during the growth. Moreover, the contrast within (Mg,Zn)O barriers exhibits segregation-induced features. Such features appear at the bottom and top groove edges with bright and dark contrast, respectively. This has been previously assessed by both STEM-based Energy Dispersive Spectroscopy (EDS) and by APT itself [18]. As it is common practice, the distances along the growth axis obtained by STEM have been subsequently used as a reference frame in order to obtain the most accurate APT reconstruction.

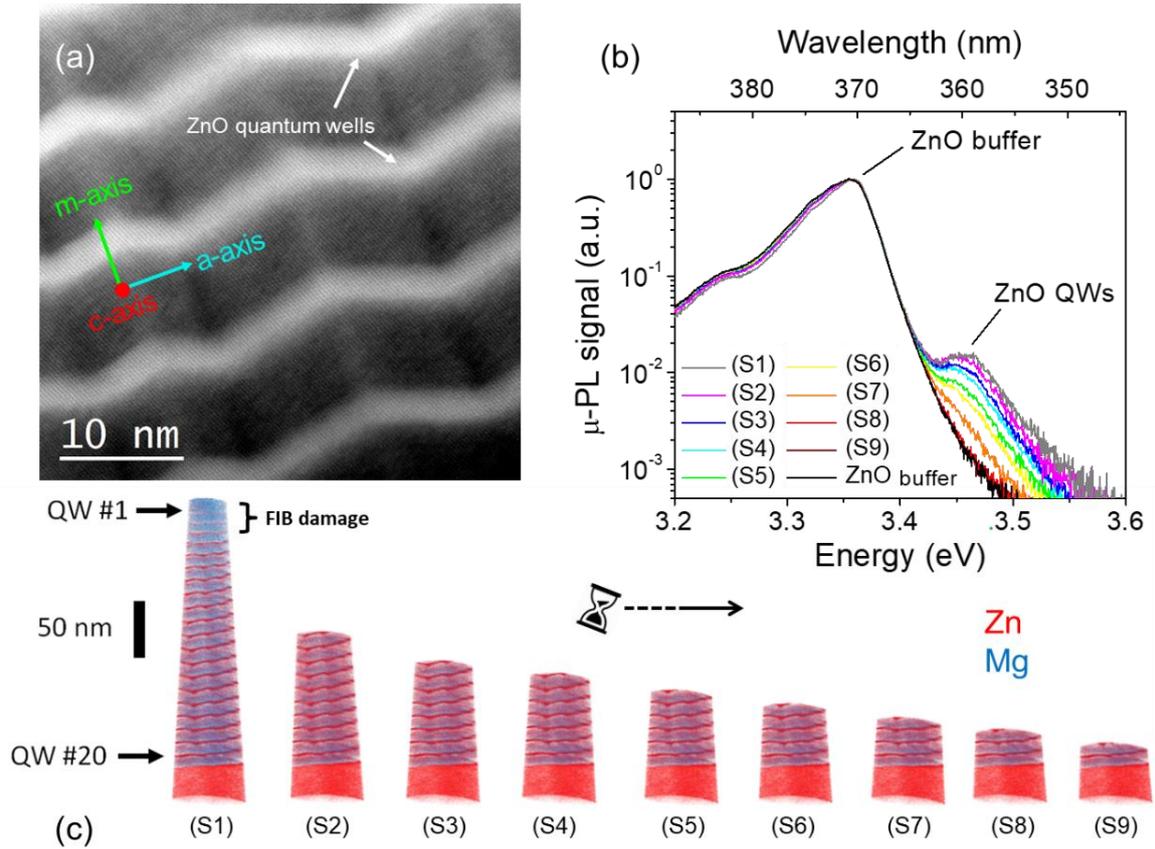

Fig. 2. (a) Transmission electron micrograph showing the ZnO/(Mg,Zn)O multi-quantum well system analyzed by simultaneous µPL and APT; (b) Sequence of µPL spectra recorded from the evaporating specimen. (c) Portions of reconstructed APT volume corresponding to the portion of the tip not yet evaporated after the acquisition of the correspondingly indexed spectrum in (b).

From this system, tip specimens have been fabricated by standard focused ion beam (FIB) process and analyzed simultaneously by µPL and APT.

### 3.1. Performance of the APT instrument

The data relative to the APT analysis are shown in Fig. 2-(b-c) and in Fig. 3. The experimental parameters have been set as follows: specimen base temperature $T_{base}$ = 80 K, laser repetition rate 500 kHz, laser pulse energy $E_{las}$ = 0.25 nJ, detection rate $\phi$ = 0.0035 event/pulse, spectrometer grating 1200 lines/mm, CCD integration time 30 s. In part (ii) we report the sequence of spectra, each of which is labeled with a progressive index. The 3D reconstructed volume extracted from the APT data is reported in Fig.2-(c). The volume has been reconstructed with a cone-angle protocol, considering the geometrical parameters of the tip specimen observed by SEM and adjusting them in order to match the interlayer distance assessed by TEM. The mass spectrum recorded during the analysis is reported in Fig. 3. The main species usually found in the analysis of MgZnO are found, such as $Mg^{2+}$, $Mg^+$, $O^+$, $O_2^+$, $Zn^{2+}$, $Zn^+$ and several types of molecular ions containing both Zn and O [18]. The composition measurement in the MgZnO barrier layers is also compatible with previous experiments performed in a LaWATAP instrument, yielding an average Mg II-site fraction of $y = 0.32$.

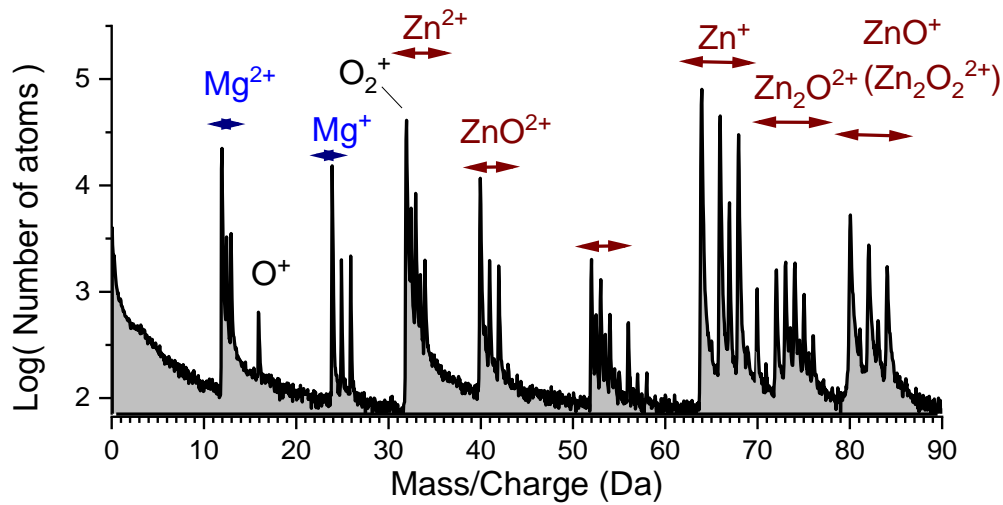

Fig. 3. Mass spectrum recorded during the analysis of the ZnO/MgZnO multi quantum well specimen.

### 3.2. Time-integrated PL and correlation with APT data

The sequence of µPL spectra collected from the tip specimen during its evaporation is reported in Fig. 2-(b). The spectra are indexed from S1 to S9. Two main spectral components can be identified. A lower-energy component, peaked at $E_{sub} = 3.375$ eV, corresponding to the emission from the ZnO substrate, and a higher-energy component, peaked at $E_{QW} \approx 3.46$ eV, corresponding to the emission from the QWs. The reconstructed volume corresponding to the portion of tip that remains to evaporate after the acquisition of each spectrum is reported in Fig. 2-(c). From the evolution of spectra and evaporated volume the correlation between the decrease of the QW signal and the evaporation of the QWs appears with clarity. It should be noticed that during the initial part of the evaporation corresponding to the topmost 8 QWs, the QW signal does not evolve significantly. This can be caused by the degradation of the topmost layers by the FIB preparation. This is even visible in the reconstructed volume corresponding to the 4 topmost QWs. On the other hand, almost no change occurs to the ZnO emission during the whole evaporation process, only a slight peak broadening on the low-energy side that could be a consequence of a slight modification of the strain state within the substrate due to the modification of the tip shape during the evaporation. As a first important consequence of this correlation, it is clear that the technique allows unambiguous attribution of either peak in the PL spectrum.

In agreement with our previous study, the QW morphological parameters indicate a thickness of $t = (4.0 \pm 0.25)$ nm, while the Mg II-site fraction in the barrier in in the range 0.32 (value corresponding to the region where the V-groove points towards the growth direction) and 0.15 (where the V-groove points towards the substrate). These compositional and morphologic features can be shown to yield an expected radiative recombination energy et $E_{QW} = 3.46$ eV, as reported in the PL spectra of Fig.-2-(b).

4. **In-situ time-resolved photoluminescence under static conditions**

In this section we provide an example of application of a time-resolved PL (TRPL) measurement within the PAP instrument under static conditions, i.e. without applied bias. The feasibility of such measurements was studied in a previous work [17], but the measurement was performed in an independent optical bench at the pulse frequency of 4MHz, which is not compatible with APT operation. The analyzed specimen is a field emission tip extracted from a ZnO/(Mg,Zn)O system containing only one optically active quantum well surrounded by (Mg,Zn)O barriers in which the average Mg II-site fraction is around 0.3. The system was epitaxially grown on a ZnO m-plane substrate. The acquired data are reported in fig. 4. The data were acquired at a specimen base temperature $T_{base}$ = 80 K, laser repetition rate 400 kHz, laser pulse energy $E_{las}$ = 0.25 nJ, spectrometer grating 150 lines/mm, streak camera integration time 600 s. Notice that these conditions are compatible with APT analysis, though the integration time in this specific example is ten times higher than that relative to the time-integrated acquisition illustrated in the previous sections. Figure 4-(a) shows the streak camera 2D histogram with the photon wavelength on the abscissa and the time delay on the ordinates. The time-integration of this dataset yields the spectrum shown in fig. 4-(b): here, the spectral contributions corresponding to ZnO (peak at $E_{sub}$=3.3 eV), to the ZnO QW (peak at $E_{QW}$=3.4 eV) and to the (Mg,Zn)O barrier (peak at $E_{barr}$=3.85 eV). The time decay transients of the PL for each peak can then be extracted, and are shown in fig. 4-(c). Significant differences may be found in the temporal decay of each component, which will be qualitatively discussed (more detailed discussions and quantitative interpretations being beyond the scope of the present work). In this specific dataset, the ZnO peak is the faintest, and also exhibits a very short transient. This is possibly due a combination of loosely-localized exciton levels and of FIB-induced damage at the specimen surface. The time decay is then governed by non-radiative recombination at defects created either at the surface or in the proximity (several tens of nm [14]) of it. The QW peak is relatively intense and decays with a transient with a time constant in the range of 0.5-0.7 ns. Excitons within the QW are supposed to be more localized at interface roughness and thus more protected from the defects on the surface. However, the significant overlap of electron and hole states may lead to an increase in the radiative recombination probability, resulting in the observed time behavior. Finally, the barrier peak, quite intense, exhibits a slower, double exponential transient with time constants in the range of 2-3 ns. The longer transient may be explained by the strong localization of electron-hole pairs within the alloy, due to alloy fluctuations: carriers are localized and thus protected from the surface defects, but the electron-hole overlap is reduced with respect to the situation in the QW, resulting in a relatively long transient.

These preliminary results show that in-situ, operando TRPL within the PAP instrument should become possible in a next future.

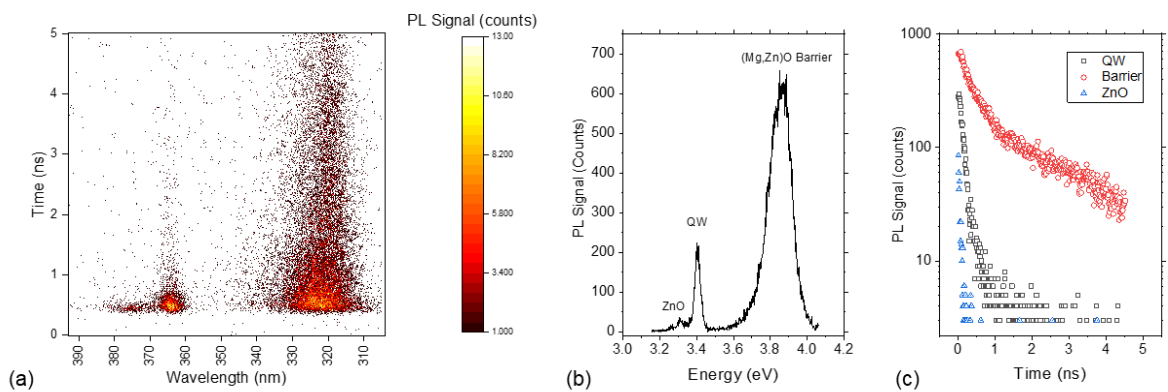

Fig. 4. Measurement of TRPL from an unbiased field-emission tip specimen containing a ZnO-(Mg,Zn)O QW system. (a) Streak camera image, (b) resulting PL spectrum and (c) PL transients extracted from the QW (black squqres), (Mg,Zn)O barrier (red circles) and ZnO (blue triangles) peaks.

## 5. Conclusions and perspectives

A coupled in-situ µPL bench within a Laser-assisted Tomographic Atom Probe has been developed. The instrument can perform simultaneous µPL spectroscopy and APT analysis at a pulse frequency of 500 kHz, with the specimen set at typical temperatures for APT analysis (20 – 80 K). As an example of its application, we have reported the analysis of a ZnO/(Mg,Zn)O multi-quantum well structure. The correlative study of optical and APT data makes it possible to attribute spectral signatures to individual quantum wells and to interpret the optical data through the analysis of the 3D morphology of the well as obtained by the APT composition maps. Furthermore, we have demonstrated that TRPL can be measured under experimental conditions similar to those in which APT analysis can be performed.

In perspective, the instrument can be applied to a large variety of materials systems, provided they contain optically active emitters. In perspective, the optical signature profiling can be applied to the study of APT specimens containing more light emitters, allowing for the attribution of each signature to a specific emitter with a spatial resolution beyond the diffraction limit of the excitation laser light (which is also referred to as *super-resolution spectroscopy*). In the case of the study of single emitters, it will be possible to consider the evolution of the optical signature of an individual radiative center during the APT analysis. In particular, we will address how the PL intensity, the peak wavelength, the time decay constant and other optical parameters evolve in correlation with the evaporation of the specimen volume. Furthermore, it will be possible to explore the idea of using single quantum emitters in APT tip specimens as optical probes close to the apex of the tip for the study of the field intensity, field-induced strain and amount of absorbed light during the laser pulse, well beyond what we could demonstrate in our exploratory study on NV centers in (non-evaporating) diamond [13]. TRPL could also help interpret the temperature evolution within the tip during and after the laser pulse. Spectral signatures could be indeed put in relationship with the variation of the material bandgap as a function of laser power or of the position of a localized light emitter through the so-called Varshni shift [19]. The knowledge of these quantities, which are estimated but were never measured, will make it possible to understand the mechanisms of laser-assisted field ion evaporation [20]. In perspective, it will be possible to use this information in continuum and atomistic models describing surface field distribution, atomic and molecular evaporation.

Finally, from the instrumental point of view, it will be straightforward to upgrade the present prototype to a higher-frequency, higher-efficiency and larger-field of view instrument, setting it at the basis of an original competitive approach in nanoscale materials science.

## ACKNOWLEDGEMENTS

This work was funded by the French National Research Agency (ANR) in the framework of the projects EMC3 Labex AQURATE, EMC3 Labex IDEPOP and ANR-13-JS10-0001-01 TAPOTER and co-funded in the framework of RIN IFROST, EMC3 Labex IDEPOP and CPER BRIDGE projects by European Union with European Regional Development Fund (ERDF) and by Region Normandie.

**AIP Publishing Data Sharing Policy**

Data available on request from the authors